\documentclass[twocolumn,showpacs,prl]{revtex4}

\usepackage{graphics,graphicx}

\newcommand{\beq}{\begin{equation}}
\newcommand{\eeq}{\end{equation}}
\newcommand{\bea}{\begin{eqnarray}}
\newcommand{\eea}{\end{eqnarray}}

\begin{document}

\title{Total recoil:  the maximum kick from nonspinning black-hole binary inspiral}

\author{
Jos\'e A. Gonz\'alez, Ulrich Sperhake, Bernd Br\"ugmann, Mark Hannam, Sascha Husa}

\affiliation{Theoretical Physics Institute, University of Jena, 07743 Jena, Germany}

\date{\today}

\begin{abstract}
When unequal-mass black holes merge, the final black hole receives a
``kick'' due to the asymmetric loss of linear momentum in the
gravitational radiation emitted during the merger. The magnitude of
this kick has important astrophysical consequences. Recent
breakthroughs in numerical relativity allow us to perform the largest
parameter study undertaken to date in numerical simulations of binary black hole
inspirals. We study non-spinning black-hole binaries with mass ratios
from $q=M_1/M_2=1$ to $q =0.25$ ($\eta = q/(1 + q)^2$ from 0.25 to
0.16). We accurately calculate the velocity of the kick to within
6\%, and the final spin of the black holes to within 2\%. A maximum
kick of $175.2\pm11$ km\,s$^{-1}$ is achieved for 
$\eta = 0.195 \pm 0.005$.
\end{abstract}


\maketitle

\paragraph{Introduction.---}

Anisotropic emission of gravitational waves from the coalescence of black-hole
binaries carries away linear momentum and thus imparts a recoil on the
merged hole. This recoil, often referred to as a ``kick'' or ``rocket effect'',
has important consequences for various astrophysical scenarios. 
The displacement or ejection
of black holes as a result of a black-hole merger not only leads to a population
of interstellar and intergalactic massive black holes, but also has severe
repercussions on the formation of supermassive black holes and
the structure of the host galaxies \cite{Boylan-Kolchin2004,Haiman:2004ve,
Madau2004, Merritt2004, Libeskind2005, Volonteri2005}. The demography of
massive black holes is also relevant for the expected number
of sources for the space-based gravitational-wave detector LISA.
The gravitational recoil might also manifest itself directly
in astrophysical observations, such as
the discovery of bright quasi-stellar objects without a host-galaxy
(see \cite{Haehnelt2005, Magain2005, Hoffman2006, Merritt2006}) 
and the distorted morphology of $\times$-shaped radio sources
\cite{Merritt2002, Madau2004, Merritt2004}.
Accurate recoil estimates are important for all of these astrophysical
models.

The strongest contribution to the kick is made during the plunge and merger of
the black holes. In this regime nonlinear general relativistic effects
preclude reliable analytic treatment, and even the most recent sophisticated
analytic estimates \cite{Favata:2004wz,Blanchet:2005rj,Damour:2006tr,Sopuerta:2006wj, Sopuerta:2006b}, 
which give a maximum kick varying from 50 to 500~km\,s$^{-1}$, 
carry uncertainties of 25 to 50\%. Numerical studies
using full numerical relativity are necessary to accurately determine the total recoil. 
In contrast to analytic calculations, numerical simulations contain only one
physical approximation: the initial data are not exactly equivalent to an astrophysical 
inspiral process. However, the resulting errors decrease as the black holes are placed 
further apart. If the initial separation is large enough that its effect on the estimate
of the kick or final angular momentum is minimal, then the result can be said to be 
accurate and free from any physical approximation. 

Recent breakthroughs in numerical relativity \cite{Pretorius:2005gq,Campanelli:2005dd,Baker05a,Bruegmann:2003aw}
have made possible long-term stable numerical evolutions of
binary black hole systems for several orbits through merger and ringdown
\cite{Pretorius:2006tp,Campanelli:2006gf,Baker:2006yw,Herrmann:2006ks,Sperhake:2006cy,Scheel-etal-2006:dual-frame,Buonanno:2006n}.
Drastic improvements in computational efficiency achieved by a new
generation of accurate
finite-difference mesh-refinement codes, e.g.\ 
\cite{Sperhake:2006cy,Bruegmann:2006at}, pave the way for 
the large parameter studies necessary to explore the parameter
space of binary black hole inspiral. Here we present the first such study, 
comprising roughly three dozen unequal-mass initial-data sets, to compute the
gravitational recoil and the spin of the final black hole. 
Implications for gravitational-wave data analysis will be explored elsewhere.

Early numerical estimates of the recoil suffered from low numerical accuracy,
or initial black-hole separations that were too small. (Compare, for example,
the value of 240$\pm$140~km\,s$^{-1}$ for $q = 0.5$ in
\cite{Campanelli:2005aa} and the estimate 33~km\,s$^{-1}$ reported in 
\cite{Herrmann:2006ks} for $q = 0.85$.)
The first accurate numerical result was recently given by Baker, {\it
et al.}~\cite{Baker:2006nr}, who found a value of
$101\pm15$~km\,s$^{-1}$ for $q = 0.67$.

In this paper we present results from numerical simulations of unequal-mass nonspinning
black-hole binaries with mass ratios $q=1.0$ to $q = 0.253$. We estimate that
the total error in the kick velocities that we quote is less than 6\% (see
below). We are thus able to improve on the numerical and physical
accuracy achieved in \cite{Baker:2006nr}, and, more importantly, our numerical
simulations for the first time cover a
range of mass ratios large enough to accurately determine the maximum kick resulting
from nonspinning binaries. We calculate the maximum kick to be 175.2 $\pm$ 11~km\,s$^{-1}$.


\paragraph{Numerical methods.---}

Numerical simulations were performed with the BAM code \cite{Bruegmann:2006at,Bruegmann:2003aw}, 
in which we have implemented the ``moving puncture'' method of evolving black-hole
binaries \cite{Campanelli:2005dd,Baker05a,Hannam:2006vv}.
This approach is based on initial data of puncture type
\cite{Bowen80,Brandt97b}, which are evolved using the BSSN/(3+1) formulation of the
Einstein equations \cite{Shibata95,Baumgarte99}, with a suitable
gauge choice. The code uses a box-based mesh-refinement grid structure with
coarser refinement levels being centered on the origin and the high resolution
levels consisting of two components centered around each hole and following
their motion across the computational domain.
The evolution uses a fourth-order accurate Runge-Kutta
integrator, and Berger-Oliger timestepping for the mesh refinement.
Gravitational waves are extracted in the form of the Newman-Penrose scalar
$\Psi_4$ on spheres of constant coordinate radius.
Details of all aspects of the implementation have been presented in
\cite{Bruegmann:2006at}.

In order to determine the physical parameters of the initial data,
we estimate the initial momenta of the black holes using the 3PN-accurate 
formula given in Sec.\ VII in
\cite{Bruegmann:2006at}, for given masses and coordinate separation. The black
holes have masses $M_1$ and $M_2$, and the total black-hole mass is $M = M_1 +
M_2$. The total gravitational energy of the system is $M_{ADM}$ \cite{Arnowitt62}. 
In the notation of \cite{Bruegmann:2006at}, we use the $\chi_{\eta=2}$ method,
and label our runs by the number of gridpoints, $i$, on the finest level. 
We have observed that grid sizes of at least
$i=48$ are required to achieve fourth-order convergence in the waveforms and 
puncture trajectories. Here we have performed runs with $i= 32,40,48,56$ for 
all mass ratios, corresponding to finest resolutions of $1/26$, $1/32$, $1/38$ 
and $1/45$ (for details see Table I in \cite{Bruegmann:2006at}). 
The main purpose of the low resolution runs was
to develop and test our strategy for setting up our simulations.
In order to confirm that our $i=56$ runs are in the fourth-order 
convergent regime for the waveforms and puncture tracks,
we have performed convergence tests using resolutions characterized by 
$i=56,64,72,80$ (with resolutions at the punctures of $1/45$, $1/51$, $1/58$,
and $1/64$) for selected mass ratios.
We start with a simulation of an equal-mass binary, that is, each black hole has an initial mass 
of $M_1=M_2=0.5$. The mass ratio is then changed by increasing
the mass of one of the holes. For example, the black holes have
masses $M_1=0.5$ and $M_2=1.0$ in the case of a mass-ratio $q=M_1/M_2=0.5$. 
The idea behind this strategy is to keep constant
the effective numerical resolution of the small
black hole, while increasing that for the larger hole. For
consistency, it is then necessary to proportionately increase the initial distance
between the black holes, the distance of the outer boundary of the numerical grid, 
and the extraction radii of the gravitational waves. We find that sufficiently 
accurate (for the purpose of this study) gravitational-wave signals could be extracted at 30$M$.
Finally, the radiated linear momentum is calculated from $\Psi_4$ according to
\beq
\frac{dP_i}{dt} = \lim_{r \rightarrow \infty} \left[ \frac{r^2}{16\pi}
  \int_{\Omega} \ell_i \left| \int_{-\infty}^{t} \Psi_4 d\tilde{t}
  \right|^2 d{\Omega}\right]. 
  \label{eqn:Pint}
\eeq
where $ \ell_i = \left(\sin \theta \cos \phi,\,\,\,\sin \theta \sin \phi,\,\,\,
\cos \theta \right)$ \cite{Campanelli99,Baker:2006nr}.

The puncture initial data with Bowen-York extrinsic 
curvature used in our simulations are known to contain spurious
gravitational radiation. This pulse quickly leaves the system, but
 carries away linear momentum which we find to
be in the direction {\it transverse} to the black-hole motion. This
small initial 
``kick'' of the order of $10\,$km\,s$^{-1}$ is not part of the astrophysical 
black-hole recoil we wish to measure. We therefore wait for the initial pulse
to pass through the extraction sphere (after about 50$\,M_{ADM}$ in our
simulations), and calculate $P_i$ by starting the integration 
of $dP_i/dt$ as given in Eq.~(\ref{eqn:Pint}) at $t_0 = 50M_{ADM}$.
In the course of the inspiral, $dP_i/dt$ oscillates around zero, and,
thus, the final integrated kick $P_i$ will depend on the choice of
$t_0$. For the separations we consider ambiguities in this choice introduce an
uncertainty in $P_i$ of 3\%.  

\paragraph{Results.---}

\begin{figure}[ht]
\centerline{\resizebox{9cm}{!}{\includegraphics[angle=-90]{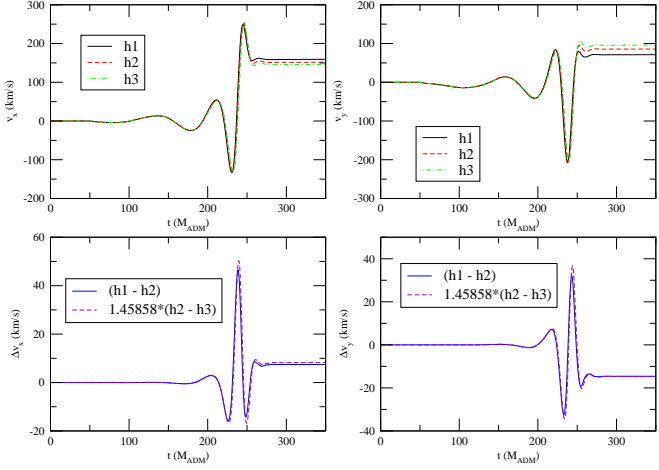}}}
\caption{Components $v_x$ and $v_y$ of the kick velocity as function of time, 
for $\eta=0.19$, for resolutions $h_1 = 1/45$, $h_2 = 1/51$ and $h_3 = 1/58$.
Top panels: kick components $v_x$ and $v_y$. Lower panels: demonstration of
second-order convergence.}
\label{fig:convergence}
\end{figure}

We performed two sets of runs. In the first set, the mass ratio was varied
from $q = 1.0$ to $q = 0.25$, and the initial coordinate
separation of the black holes was kept fixed at $r_0 = 7.0M$. 
Convergence tests were performed for mass ratios $q=0.4, 0.33, 0.286$
 with finest-grid resolutions of $h_1 = 1/45$, $h_2 = 1/51$ and $h_3 = 1/58$. 
In order to test the convergence properties of the recoil,
we study the components $P_x$ and $P_y$ of the final kick.
We find the results
to be consistent with second-order convergence, as shown for 
$q = 0.33$ in Fig.~\ref{fig:convergence}.
The kick is calculated by integrating twice in time a function that is several
orders of magnitude smaller than the wave signal $\Psi_4 (2,2)$, and we
believe this to be the reason for the lower-order accuracy than observed in
\cite{Bruegmann:2006at}. Higher resolutions, currently beyond our computational
resources, are necessary to achieve
the limiting overall convergence behavior of the code for all quantities. 
There is a large error in the estimate of the kick's direction, but the kick's
magnitude is calculated with high accuracy. We estimate that the total kicks
are calculated with a numerical error of less than 2\%.

From Fig.~\ref{fig:convergence} we see that $v_x=P_x/M_{F}$ and $v_y=P_y/M_{F}$, 
where $M_F$ is the mass of the final black hole, oscillate around zero during the 
inspiral. This is consistent with a small continuous loss of linear momentum, 
like water from a spinning lawn sprinkler, which pushes the center-of-mass in a 
roughly circular motion, before the final kick during merger.

For the second set of runs we fix $\eta=0.19$, but
vary the initial separations, setting $r_0 = 6.0M$, $7.0M$ and $8.0M$. 
These simulations demonstrate the extra contribution to the kick at 
larger $r_0$ to be small, so that we estimate the accumulated kick from the 
early inspiral to be less than 1\%. This is comparable to post-Newtonian
estimates of the contribution to the kick up to our 
initial separation \cite{Blanchet:2005rj}.

\begin{figure}[ht]
\centerline{
\includegraphics[width=6.8cm,height=9cm,angle=-90]{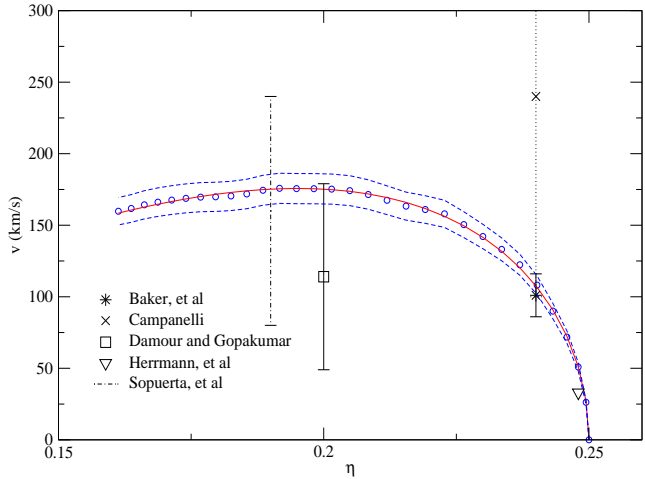}}
\caption{The kick velocity as a function of mass ratio, with an error of
  $\pm6$\% indicated by the dotted lines.
We also indicate previous numerical results from Baker, {\it et al}
\cite{Baker:2006nr}, Campanelli \cite{Campanelli:2005aa}, and Herrmann, {\it et al}
\cite{Herrmann:2006ks}, and the analytic estimates of Damour and Gopakumar
\cite{Damour:2006tr} and Sopuerta, {\it et al} \cite{Sopuerta:2006wj}.}
\label{fig:kicks}
\end{figure}

\begin{figure}[ht]
\centerline{\resizebox{9cm}{!}{\includegraphics[angle=-90]{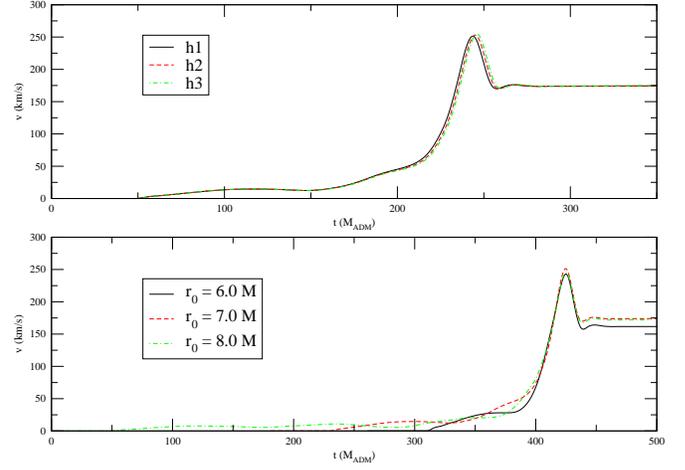}}}
\caption{Total kick velocity ($v=\sqrt{v_x^2+v_y^2}$) as function of time
for  $\eta=0.19$.
Top panel: for resolutions $h_1, h_2$ and $h_3$ as described
in the text. Lower panel: for runs with three initial black-hole
separations $r_0 = 6.0M, 7.0M, 8.0M$.}
\label{fig:kick_vs_time}
\end{figure}

Figure~\ref{fig:kicks} shows the total kick from all of these runs, as
a function of $\eta$. Combining all the errors we have discussed (3\% due to
the choice of $t_0$ in Eqn.~(\ref{eqn:Pint}), 2\% numerical error, and 1\% due
to the neglected contribution from earlier inspiral), we
conservatively estimate a total error of less than 6\%. 
A least-squares fit of the kick with $v = A \eta^2
\sqrt{1 - 4 \eta} (1 + B\eta)$ (based on the formula of Fitchett
\cite{Fitchett83})  gives $A = 1.20\times10^4$ and $B = -0.93$ with
$\chi^2 = 48.25$ with $\nu=29$ degrees of freedom. 
From our curve fit we calculate a maximum kick of
$V_{max} = 175.2\pm 11$km\,s$^{-1}$ at $\eta = 0.195\pm 0.005$ ($q = 0.36\pm
0.03$). This agrees with the estimate of 114 $\pm$ 65~{\rm km\,s}$^{-1}$
of \cite{Damour:2006tr}, and also the
close-limit analyses in~\cite{Sopuerta:2006wj, Sopuerta:2006b}. The higher
estimate of $250\pm 50\,{\rm km/s}$ reported in Ref.\,\cite{Blanchet:2005rj}
does not include a possible ``breaking'' effect in the ringdown phase, and
their values agree well with the local maximum in Fig.~\ref{fig:kick_vs_time}.

\begin{figure}[ht]
\centerline{
\includegraphics[width=5cm,height=8.8cm,angle=-90]{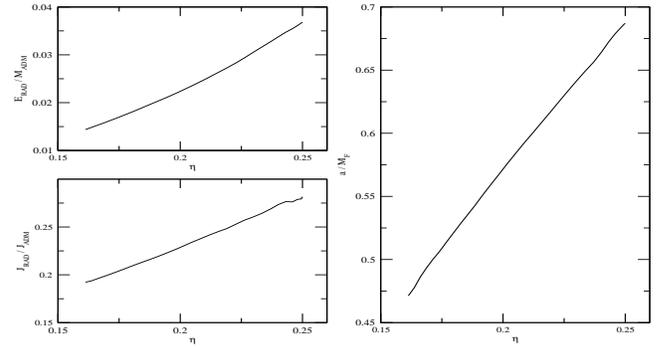} }
\caption{
Left panel: Radiated energy and angular momentum as function of the mass ratio.
Right panel: 
Spin of the final black hole ($a/M_f$) as function of the mass ratio.
The spin curves can be fit by $a/M_f = 0.089 (\pm 0.003) + 2.4 (\pm 0.025) \eta$.}
\label{fig:E_and_J_vs_eta}
\end{figure}

Finally, we address the spin of the merged black hole.
An understanding of the demographics of black holes, in particular the expected values of spins, 
is of essential importance for astrophysics, and
also for developing approaches to 
explore the astrophysically relevant 
binary black hole inspiral parameter space. 
We have computed the initial  angular momentum from surface
integrals at the wave extraction sphere as described in \cite{Bruegmann:2006at}, 
and the final  angular momentum from computing the wave ringdown frequency 
from an amplitude-phase decomposition of the radiation signal, and comparing with the
dependence of angular momentum of a Kerr black hole on the ringdown frequency
as quoted in \cite{Berti06b}. 
An error estimate is obtained from evaluating the angular momentum surface integrals
at the extraction sphere at the end of the simulation. We thus find our results to be
accurate to within about $2\,\%$.
Our results for the unequal-mass sequence considered
here are displayed in Fig.~\ref{fig:E_and_J_vs_eta}. In the regime we consider, the dependence
of the final spin on the mass ratio is approximately linear when expressed as a function of $\eta$:
$a/M_f = 0.089 (\pm 0.003) + 2.4 (\pm 0.025) \eta$ (the correct result for $\eta=0$ corresponds to
$a=0$).

\paragraph{Discussion.---}

Our results constitute the first accurate numerical calculations of the recoil
velocity from the merger of nonspinning black holes for a large range of mass
ratios, $q = 1.0$ to $0.253$. This range allows us to determine the maximum
kick as $V_{max} = 175.2\pm 11$~km\,s$^{-1}$ at a mass ratio of $q = 0.36\pm 0.03$.
Theoretical estimates predict a maximum kick at $q = 0.38$, and agree
well with our value. Our maximum kick velocity is within the
error bounds of the analytic estimates of Damour and Gopakumar
\cite{Damour:2006tr} and Sopuerta, {\it et al} \cite{Sopuerta:2006wj}, which
have significantly refined an earlier estimate by Favata {\it et al}
\cite{Favata:2004wz}. The kick before ringdown agrees well
with the results in \cite{Blanchet:2005rj}.
We also find consistency with the numerically calculated value $105\pm10$~km\,s$^{-1}$
for $q = 0.67$ in \cite{Baker:2006nr}.

The results presented here required 
the largest parameter study of numerical 
binary black hole evolutions to date, comprising approximately three dozen data sets 
describing unequal-mass 
nonspinning inspiraling black holes. The efficiency of our code \cite{Bruegmann:2006at} 
has enabled us to perform all the simulations quoted here at 
a total computational cost 
(including the development of our strategy and setup, production runs
and convergence tests) of about 150 000 CPU hours. 

In future work we plan to study the consequences of our results for
gravitational-wave data analysis.
Extending this work to much higher mass ratios would require 
larger evolution times and higher resolutions, making such simulations
computationally very expensive, but in principle possible with our current
techniques. The situation is similar if we wish to consider larger initial
separations.


\acknowledgments
We thank Emanuele Berti for giving us access to data
for the quasinormal mode frequencies of Kerr spacetime, Clifford Will and Luc
Blanchet for discussions of the ringdown breaking, and
Iris Christadler at the Leibniz Computer
Center (LRZ) for assistance in code optimization.
Computations where performed at HLRS (Stuttgart) and LRZ (Munich).
This work was supported in part by
DFG grant SFB/Transregio~7 ``Gravitational Wave Astronomy''.

\bibliography{references_cvs,references_extra}

\begin{thebibliography}{39}
\expandafter\ifx\csname natexlab\endcsname\relax\def\natexlab#1{#1}\fi
\expandafter\ifx\csname bibnamefont\endcsname\relax
  \def\bibnamefont#1{#1}\fi
\expandafter\ifx\csname bibfnamefont\endcsname\relax
  \def\bibfnamefont#1{#1}\fi
\expandafter\ifx\csname citenamefont\endcsname\relax
  \def\citenamefont#1{#1}\fi
\expandafter\ifx\csname url\endcsname\relax
  \def\url#1{\texttt{#1}}\fi
\expandafter\ifx\csname urlprefix\endcsname\relax\def\urlprefix{URL }\fi
\providecommand{\bibinfo}[2]{#2}
\providecommand{\eprint}[2][]{\url{#2}}

\bibitem[{\citenamefont{Boylan-Kolchin
  et~al.}(2004)\citenamefont{Boylan-Kolchin, Ma, and
  Quataert}}]{Boylan-Kolchin2004}
\bibinfo{author}{\bibfnamefont{M.}~\bibnamefont{Boylan-Kolchin}},
  \bibinfo{author}{\bibfnamefont{C.-P.} \bibnamefont{Ma}}, \bibnamefont{and}
  \bibinfo{author}{\bibfnamefont{E.}~\bibnamefont{Quataert}},
  \bibinfo{journal}{Astrophys. J.} \textbf{\bibinfo{volume}{613}},
  \bibinfo{pages}{L37} (\bibinfo{year}{2004}).

\bibitem[{\citenamefont{Madau and Quataet}(2004)}]{Madau2004}
\bibinfo{author}{\bibfnamefont{P.}~\bibnamefont{Madau}} \bibnamefont{and}
  \bibinfo{author}{\bibfnamefont{E.}~\bibnamefont{Quataet}},
  \bibinfo{journal}{Astrophys.J} \textbf{\bibinfo{volume}{606}},
  \bibinfo{pages}{L17} (\bibinfo{year}{2004}).

\bibitem[{\citenamefont{Merritt et~al.}(2004)\citenamefont{Merritt,
  Milosavljevi{\'c}, Favata, Hughes, and Holz}}]{Merritt2004}
\bibinfo{author}{\bibfnamefont{D.}~\bibnamefont{Merritt}},
  \bibinfo{author}{\bibfnamefont{M.}~\bibnamefont{Milosavljevi{\'c}}},
  \bibinfo{author}{\bibfnamefont{M.}~\bibnamefont{Favata}},
  \bibinfo{author}{\bibfnamefont{S.~A.} \bibnamefont{Hughes}},
  \bibnamefont{and} \bibinfo{author}{\bibfnamefont{D.~E.} \bibnamefont{Holz}},
  \bibinfo{journal}{Astrophys. J.} \textbf{\bibinfo{volume}{607}},
  \bibinfo{pages}{L9} (\bibinfo{year}{2004}).

\bibitem[{\citenamefont{Libeskind et~al.}(2006)\citenamefont{Libeskind, Cole,
  Frenk, and Helly}}]{Libeskind2005}
\bibinfo{author}{\bibfnamefont{N.~I.} \bibnamefont{Libeskind}},
  \bibinfo{author}{\bibfnamefont{S.}~\bibnamefont{Cole}},
  \bibinfo{author}{\bibfnamefont{C.~S.} \bibnamefont{Frenk}}, \bibnamefont{and}
  \bibinfo{author}{\bibfnamefont{J.~C.} \bibnamefont{Helly}},
  \bibinfo{journal}{MNRAS} \textbf{\bibinfo{volume}{368}},
  \bibinfo{pages}{1381} (\bibinfo{year}{2006}).

\bibitem[{\citenamefont{Volonteri and Perna}(2005)}]{Volonteri2005}
\bibinfo{author}{\bibfnamefont{M.}~\bibnamefont{Volonteri}} \bibnamefont{and}
  \bibinfo{author}{\bibfnamefont{R.}~\bibnamefont{Perna}},
  \bibinfo{journal}{MNRAS} \textbf{\bibinfo{volume}{358}}, \bibinfo{pages}{913}
  (\bibinfo{year}{2005}).

\bibitem[{\citenamefont{Haiman}(2004)}]{Haiman:2004ve}
\bibinfo{author}{\bibfnamefont{Z.}~\bibnamefont{Haiman}},
  \bibinfo{journal}{Astrophys. J.} \textbf{\bibinfo{volume}{613}},
  \bibinfo{pages}{36} (\bibinfo{year}{2004}).

\bibitem[{\citenamefont{Haehnelt et~al.}(2006)\citenamefont{Haehnelt, Davies,
  and Rees}}]{Haehnelt2005}
\bibinfo{author}{\bibfnamefont{M.~G.} \bibnamefont{Haehnelt}},
  \bibinfo{author}{\bibfnamefont{M.~B.} \bibnamefont{Davies}},
  \bibnamefont{and} \bibinfo{author}{\bibfnamefont{M.~J.} \bibnamefont{Rees}},
  \bibinfo{journal}{MNRAS} \textbf{\bibinfo{volume}{366}}, \bibinfo{pages}{L22}
  (\bibinfo{year}{2006}).

\bibitem[{\citenamefont{Magain et~al.}(2005)\citenamefont{Magain, Letawe,
  Courbin, Jablonka, Jahnke, Meylan, and Wisotzki}}]{Magain2005}
\bibinfo{author}{\bibfnamefont{P.}~\bibnamefont{Magain}},
  \bibinfo{author}{\bibfnamefont{G.}~\bibnamefont{Letawe}},
  \bibinfo{author}{\bibfnamefont{F.}~\bibnamefont{Courbin}},
  \bibinfo{author}{\bibfnamefont{P.}~\bibnamefont{Jablonka}},
  \bibinfo{author}{\bibfnamefont{K.}~\bibnamefont{Jahnke}},
  \bibinfo{author}{\bibfnamefont{G.}~\bibnamefont{Meylan}}, \bibnamefont{and}
  \bibinfo{author}{\bibfnamefont{L.}~\bibnamefont{Wisotzki}},
  \bibinfo{journal}{Nature} \textbf{\bibinfo{volume}{437}},
  \bibinfo{pages}{381} (\bibinfo{year}{2005}).

\bibitem[{\citenamefont{Hoffman and Loeb}(2006)}]{Hoffman2006}
\bibinfo{author}{\bibfnamefont{L.}~\bibnamefont{Hoffman}} \bibnamefont{and}
  \bibinfo{author}{\bibfnamefont{A.}~\bibnamefont{Loeb}},
  \bibinfo{journal}{Astrophys. J.} \textbf{\bibinfo{volume}{638}},
  \bibinfo{pages}{L75} (\bibinfo{year}{2006}).

\bibitem[{\citenamefont{Merritt et~al.}(2006)\citenamefont{Merritt,
  Storchi-Bergmann, Robinson, Batcheldor, Axon, and Fernandes}}]{Merritt2006}
\bibinfo{author}{\bibfnamefont{D.}~\bibnamefont{Merritt}},
  \bibinfo{author}{\bibfnamefont{T.}~\bibnamefont{Storchi-Bergmann}},
  \bibinfo{author}{\bibfnamefont{A.}~\bibnamefont{Robinson}},
  \bibinfo{author}{\bibfnamefont{D.}~\bibnamefont{Batcheldor}},
  \bibinfo{author}{\bibfnamefont{D.}~\bibnamefont{Axon}}, \bibnamefont{and}
  \bibinfo{author}{\bibfnamefont{R.~C.} \bibnamefont{Fernandes}},
  \bibinfo{journal}{MNRAS} \textbf{\bibinfo{volume}{367}},
  \bibinfo{pages}{1746} (\bibinfo{year}{2006}).

\bibitem[{\citenamefont{Merritt and Ekers}(2002)}]{Merritt2002}
\bibinfo{author}{\bibfnamefont{D.}~\bibnamefont{Merritt}} \bibnamefont{and}
  \bibinfo{author}{\bibfnamefont{R.~D.} \bibnamefont{Ekers}},
  \bibinfo{journal}{Science} \textbf{\bibinfo{volume}{297}},
  \bibinfo{pages}{1310} (\bibinfo{year}{2002}).

\bibitem[{\citenamefont{Favata et~al.}(2004)\citenamefont{Favata, Hughes, and
  Holz}}]{Favata:2004wz}
\bibinfo{author}{\bibfnamefont{M.}~\bibnamefont{Favata}},
  \bibinfo{author}{\bibfnamefont{S.~A.} \bibnamefont{Hughes}},
  \bibnamefont{and} \bibinfo{author}{\bibfnamefont{D.~E.} \bibnamefont{Holz}},
  \bibinfo{journal}{Astrophys. J.} \textbf{\bibinfo{volume}{607}},
  \bibinfo{pages}{L5} (\bibinfo{year}{2004}), \eprint{astro-ph/0402056}.

\bibitem[{\citenamefont{Blanchet et~al.}(2005)\citenamefont{Blanchet, Qusailah,
  and Will}}]{Blanchet:2005rj}
\bibinfo{author}{\bibfnamefont{L.}~\bibnamefont{Blanchet}},
  \bibinfo{author}{\bibfnamefont{M.~S.~S.} \bibnamefont{Qusailah}},
  \bibnamefont{and} \bibinfo{author}{\bibfnamefont{C.~M.} \bibnamefont{Will}},
  \bibinfo{journal}{Astrophys. J.} \textbf{\bibinfo{volume}{635}},
  \bibinfo{pages}{508} (\bibinfo{year}{2005}).

\bibitem[{\citenamefont{Damour and Gopakumar}(2006)}]{Damour:2006tr}
\bibinfo{author}{\bibfnamefont{T.}~\bibnamefont{Damour}} \bibnamefont{and}
  \bibinfo{author}{\bibfnamefont{A.}~\bibnamefont{Gopakumar}},
  \bibinfo{journal}{Phys. Rev.} \textbf{\bibinfo{volume}{D73}},
  \bibinfo{pages}{124006} (\bibinfo{year}{2006}).

\bibitem[{\citenamefont{Sopuerta
  et~al.}(2006{\natexlab{a}})\citenamefont{Sopuerta, Yunes, and
  Laguna}}]{Sopuerta:2006wj}
\bibinfo{author}{\bibfnamefont{C.~F.} \bibnamefont{Sopuerta}},
  \bibinfo{author}{\bibfnamefont{N.}~\bibnamefont{Yunes}}, \bibnamefont{and}
  \bibinfo{author}{\bibfnamefont{P.}~\bibnamefont{Laguna}}
  (\bibinfo{year}{2006}{\natexlab{a}}), \eprint{astro-ph/0608600}.

\bibitem[{\citenamefont{Sopuerta
  et~al.}(2006{\natexlab{b}})\citenamefont{Sopuerta, Yunes, and
  Laguna}}]{Sopuerta:2006b}
\bibinfo{author}{\bibfnamefont{C.~F.} \bibnamefont{Sopuerta}},
  \bibinfo{author}{\bibfnamefont{N.}~\bibnamefont{Yunes}}, \bibnamefont{and}
  \bibinfo{author}{\bibfnamefont{P.}~\bibnamefont{Laguna}}
  (\bibinfo{year}{2006}{\natexlab{b}}), \bibinfo{note}{gr-qc/0611110}.

\bibitem[{\citenamefont{Pretorius}(2005)}]{Pretorius:2005gq}
\bibinfo{author}{\bibfnamefont{F.}~\bibnamefont{Pretorius}},
  \bibinfo{journal}{Phys. Rev. Lett.} \textbf{\bibinfo{volume}{95}},
  \bibinfo{pages}{121101} (\bibinfo{year}{2005}).

\bibitem[{\citenamefont{Campanelli
  et~al.}(2006{\natexlab{a}})\citenamefont{Campanelli, Lousto, Marronetti, and
  Zlochower}}]{Campanelli:2005dd}
\bibinfo{author}{\bibfnamefont{M.}~\bibnamefont{Campanelli}},
  \bibinfo{author}{\bibfnamefont{C.~O.} \bibnamefont{Lousto}},
  \bibinfo{author}{\bibfnamefont{P.}~\bibnamefont{Marronetti}},
  \bibnamefont{and}
  \bibinfo{author}{\bibfnamefont{Y.}~\bibnamefont{Zlochower}},
  \bibinfo{journal}{Phys. Rev. Lett.} \textbf{\bibinfo{volume}{96}},
  \bibinfo{pages}{111101} (\bibinfo{year}{2006}{\natexlab{a}}).

\bibitem[{\citenamefont{Baker et~al.}(2006{\natexlab{a}})\citenamefont{Baker,
  Centrella, Choi, Koppitz, and van Meter}}]{Baker05a}
\bibinfo{author}{\bibfnamefont{J.~G.} \bibnamefont{Baker}},
  \bibinfo{author}{\bibfnamefont{J.}~\bibnamefont{Centrella}},
  \bibinfo{author}{\bibfnamefont{D.-I.} \bibnamefont{Choi}},
  \bibinfo{author}{\bibfnamefont{M.}~\bibnamefont{Koppitz}}, \bibnamefont{and}
  \bibinfo{author}{\bibfnamefont{J.}~\bibnamefont{van Meter}},
  \bibinfo{journal}{Phys. Rev. Lett.} \textbf{\bibinfo{volume}{96}},
  \bibinfo{pages}{111102} (\bibinfo{year}{2006}{\natexlab{a}}).

\bibitem[{\citenamefont{Br{\"u}gmann et~al.}(2004)\citenamefont{Br{\"u}gmann,
  Tichy, and Jansen}}]{Bruegmann:2003aw}
\bibinfo{author}{\bibfnamefont{B.}~\bibnamefont{Br{\"u}gmann}},
  \bibinfo{author}{\bibfnamefont{W.}~\bibnamefont{Tichy}}, \bibnamefont{and}
  \bibinfo{author}{\bibfnamefont{N.}~\bibnamefont{Jansen}},
  \bibinfo{journal}{Phys. Rev. Lett.} \textbf{\bibinfo{volume}{92}},
  \bibinfo{pages}{211101} (\bibinfo{year}{2004}), \eprint{gr-qc/0312112}.

\bibitem[{\citenamefont{Pretorius}(2006)}]{Pretorius:2006tp}
\bibinfo{author}{\bibfnamefont{F.}~\bibnamefont{Pretorius}},
  \bibinfo{journal}{Class. Quantum Grav.} \textbf{\bibinfo{volume}{23}},
  \bibinfo{pages}{S529} (\bibinfo{year}{2006}).

\bibitem[{\citenamefont{Campanelli
  et~al.}(2006{\natexlab{b}})\citenamefont{Campanelli, Lousto, and
  Zlochower}}]{Campanelli:2006gf}
\bibinfo{author}{\bibfnamefont{M.}~\bibnamefont{Campanelli}},
  \bibinfo{author}{\bibfnamefont{C.~O.} \bibnamefont{Lousto}},
  \bibnamefont{and}
  \bibinfo{author}{\bibfnamefont{Y.}~\bibnamefont{Zlochower}},
  \bibinfo{journal}{Phys. Rev. D} \textbf{\bibinfo{volume}{73}},
  \bibinfo{pages}{061501(R)} (\bibinfo{year}{2006}{\natexlab{b}}),
  \eprint{gr-qc/0601091}.

\bibitem[{\citenamefont{Baker et~al.}(2006{\natexlab{b}})\citenamefont{Baker,
  Centrella, Choi, Koppitz, and van Meter}}]{Baker:2006yw}
\bibinfo{author}{\bibfnamefont{J.~G.} \bibnamefont{Baker}},
  \bibinfo{author}{\bibfnamefont{J.}~\bibnamefont{Centrella}},
  \bibinfo{author}{\bibfnamefont{D.-I.} \bibnamefont{Choi}},
  \bibinfo{author}{\bibfnamefont{M.}~\bibnamefont{Koppitz}}, \bibnamefont{and}
  \bibinfo{author}{\bibfnamefont{J.}~\bibnamefont{van Meter}},
  \bibinfo{journal}{Phys. Rev. D} \textbf{\bibinfo{volume}{73}},
  \bibinfo{pages}{104002} (\bibinfo{year}{2006}{\natexlab{b}}),
  \eprint{gr-qc/0602026}.

\bibitem[{\citenamefont{Herrmann et~al.}(2006)\citenamefont{Herrmann,
  Shoemaker, and Laguna}}]{Herrmann:2006ks}
\bibinfo{author}{\bibfnamefont{F.}~\bibnamefont{Herrmann}},
  \bibinfo{author}{\bibfnamefont{D.}~\bibnamefont{Shoemaker}},
  \bibnamefont{and} \bibinfo{author}{\bibfnamefont{P.}~\bibnamefont{Laguna}}
  (\bibinfo{year}{2006}), \eprint{gr-qc/0601026}.

\bibitem[{\citenamefont{Sperhake}(2006)}]{Sperhake:2006cy}
\bibinfo{author}{\bibfnamefont{U.}~\bibnamefont{Sperhake}}
  (\bibinfo{year}{2006}), \eprint{gr-qc/0606079}.

\bibitem[{\citenamefont{Scheel et~al.}(2006)\citenamefont{Scheel, Pfeiffer,
  Lindblom, Kidder, Rinne, and Teukolsky}}]{Scheel-etal-2006:dual-frame}
\bibinfo{author}{\bibfnamefont{M.~A.} \bibnamefont{Scheel}},
  \bibinfo{author}{\bibfnamefont{H.~P.} \bibnamefont{Pfeiffer}},
  \bibinfo{author}{\bibfnamefont{L.}~\bibnamefont{Lindblom}},
  \bibinfo{author}{\bibfnamefont{L.~E.} \bibnamefont{Kidder}},
  \bibinfo{author}{\bibfnamefont{O.}~\bibnamefont{Rinne}}, \bibnamefont{and}
  \bibinfo{author}{\bibfnamefont{S.~A.} \bibnamefont{Teukolsky}}
  (\bibinfo{year}{2006}), \eprint{gr-qc/0607056}.

\bibitem[{\citenamefont{Buonanno et~al.}(2006)\citenamefont{Buonanno, Cook, and
  Pretorius}}]{Buonanno:2006n}
\bibinfo{author}{\bibfnamefont{A.}~\bibnamefont{Buonanno}},
  \bibinfo{author}{\bibfnamefont{G.}~\bibnamefont{Cook}}, \bibnamefont{and}
  \bibinfo{author}{\bibfnamefont{F.}~\bibnamefont{Pretorius}}
  (\bibinfo{year}{2006}), \bibinfo{note}{gr-qc/0610122}.

\bibitem[{\citenamefont{Br{\"u}gmann et~al.}(2006)\citenamefont{Br{\"u}gmann,
  Gonz\'alez, Hannam, Husa, Sperhake, and Tichy}}]{Bruegmann:2006at}
\bibinfo{author}{\bibfnamefont{B.}~\bibnamefont{Br{\"u}gmann}},
  \bibinfo{author}{\bibfnamefont{J.~A.} \bibnamefont{Gonz\'alez}},
  \bibinfo{author}{\bibfnamefont{M.}~\bibnamefont{Hannam}},
  \bibinfo{author}{\bibfnamefont{S.}~\bibnamefont{Husa}},
  \bibinfo{author}{\bibfnamefont{U.}~\bibnamefont{Sperhake}}, \bibnamefont{and}
  \bibinfo{author}{\bibfnamefont{W.}~\bibnamefont{Tichy}}
  (\bibinfo{year}{2006}), \eprint{gr-qc/0610128}.

\bibitem[{\citenamefont{Campanelli}(2005)}]{Campanelli:2005aa}
\bibinfo{author}{\bibfnamefont{M.}~\bibnamefont{Campanelli}},
  \bibinfo{journal}{Class.Quant.Grav} \textbf{\bibinfo{volume}{22}},
  \bibinfo{pages}{S387} (\bibinfo{year}{2005}).

\bibitem[{\citenamefont{Baker et~al.}(2006{\natexlab{c}})\citenamefont{Baker,
  Centrella, Choi, Koppitz, van Meter, and Miller}}]{Baker:2006nr}
\bibinfo{author}{\bibfnamefont{J.~G.} \bibnamefont{Baker}},
  \bibinfo{author}{\bibfnamefont{J.}~\bibnamefont{Centrella}},
  \bibinfo{author}{\bibfnamefont{D.-I.} \bibnamefont{Choi}},
  \bibinfo{author}{\bibfnamefont{M.}~\bibnamefont{Koppitz}},
  \bibinfo{author}{\bibfnamefont{J.}~\bibnamefont{van Meter}},
  \bibnamefont{and} \bibinfo{author}{\bibfnamefont{M.~C.} \bibnamefont{Miller}}
  (\bibinfo{year}{2006}{\natexlab{c}}), \eprint{astro-ph/0603204}.

\bibitem[{\citenamefont{Hannam et~al.}(2006)\citenamefont{Hannam, Husa,
  Pollney, Br{\"u}gmann, and {\'O~Murchadha}}}]{Hannam:2006vv}
\bibinfo{author}{\bibfnamefont{M.}~\bibnamefont{Hannam}},
  \bibinfo{author}{\bibfnamefont{S.}~\bibnamefont{Husa}},
  \bibinfo{author}{\bibfnamefont{D.}~\bibnamefont{Pollney}},
  \bibinfo{author}{\bibfnamefont{B.}~\bibnamefont{Br{\"u}gmann}},
  \bibnamefont{and}
  \bibinfo{author}{\bibfnamefont{N.}~\bibnamefont{{\'O~Murchadha}}}
  (\bibinfo{year}{2006}), \eprint{gr-qc/0606099}.

\bibitem[{\citenamefont{Bowen and York}(1980)}]{Bowen80}
\bibinfo{author}{\bibfnamefont{J.~M.} \bibnamefont{Bowen}} \bibnamefont{and}
  \bibinfo{author}{\bibfnamefont{J.~W.} \bibnamefont{York}},
  \bibinfo{journal}{Phys. Rev. D} \textbf{\bibinfo{volume}{21}},
  \bibinfo{pages}{2047} (\bibinfo{year}{1980}).

\bibitem[{\citenamefont{Brandt and Br{\"u}gmann}(1997)}]{Brandt97b}
\bibinfo{author}{\bibfnamefont{S.}~\bibnamefont{Brandt}} \bibnamefont{and}
  \bibinfo{author}{\bibfnamefont{B.}~\bibnamefont{Br{\"u}gmann}},
  \bibinfo{journal}{Phys. Rev. Lett.} \textbf{\bibinfo{volume}{78}},
  \bibinfo{pages}{3606} (\bibinfo{year}{1997}), \eprint{gr-qc/9703066}.

\bibitem[{\citenamefont{Shibata and Nakamura}(1995)}]{Shibata95}
\bibinfo{author}{\bibfnamefont{M.}~\bibnamefont{Shibata}} \bibnamefont{and}
  \bibinfo{author}{\bibfnamefont{T.}~\bibnamefont{Nakamura}},
  \bibinfo{journal}{Phys. Rev. D} \textbf{\bibinfo{volume}{52}},
  \bibinfo{pages}{5428} (\bibinfo{year}{1995}).

\bibitem[{\citenamefont{Baumgarte and Shapiro}(1999)}]{Baumgarte99}
\bibinfo{author}{\bibfnamefont{T.~W.} \bibnamefont{Baumgarte}}
  \bibnamefont{and} \bibinfo{author}{\bibfnamefont{S.~L.}
  \bibnamefont{Shapiro}}, \bibinfo{journal}{Phys. Rev. D}
  \textbf{\bibinfo{volume}{59}}, \bibinfo{pages}{024007}
  (\bibinfo{year}{1999}).

\bibitem[{\citenamefont{Arnowitt et~al.}(1962)\citenamefont{Arnowitt, Deser,
  and Misner}}]{Arnowitt62}
\bibinfo{author}{\bibfnamefont{R.}~\bibnamefont{Arnowitt}},
  \bibinfo{author}{\bibfnamefont{S.}~\bibnamefont{Deser}}, \bibnamefont{and}
  \bibinfo{author}{\bibfnamefont{C.~W.} \bibnamefont{Misner}}, in
  \emph{\bibinfo{booktitle}{Gravitation: An introduction to current research}},
  edited by \bibinfo{editor}{\bibfnamefont{L.}~\bibnamefont{Witten}}
  (\bibinfo{publisher}{John Wiley}, \bibinfo{address}{New York},
  \bibinfo{year}{1962}), pp. \bibinfo{pages}{227--265}, \eprint{gr-qc/0405109}.

\bibitem[{\citenamefont{Campanelli and Lousto}(1999)}]{Campanelli99}
\bibinfo{author}{\bibfnamefont{M.}~\bibnamefont{Campanelli}} \bibnamefont{and}
  \bibinfo{author}{\bibfnamefont{C.~O.} \bibnamefont{Lousto}},
  \bibinfo{journal}{Phys. Rev. D} \textbf{\bibinfo{volume}{59}},
  \bibinfo{pages}{124022} (\bibinfo{year}{1999}), \eprint{gr-qc/9811019}.

\bibitem[{\citenamefont{Fitchett}(1983)}]{Fitchett83}
\bibinfo{author}{\bibfnamefont{M.}~\bibnamefont{Fitchett}},
  \bibinfo{journal}{MNRAS} \textbf{\bibinfo{volume}{203}},
  \bibinfo{pages}{1049} (\bibinfo{year}{1983}).

\bibitem[{\citenamefont{Berti et~al.}(2006)\citenamefont{Berti, Cardoso, and
  Will}}]{Berti06b}
\bibinfo{author}{\bibfnamefont{E.}~\bibnamefont{Berti}},
  \bibinfo{author}{\bibfnamefont{V.}~\bibnamefont{Cardoso}}, \bibnamefont{and}
  \bibinfo{author}{\bibfnamefont{C.~M.} \bibnamefont{Will}},
  \bibinfo{journal}{Phys. Rev. D} \textbf{\bibinfo{volume}{73}},
  \bibinfo{pages}{064030} (\bibinfo{year}{2006}), \eprint{gr-qc/0512160}.

\end{thebibliography}

\end{document}